\documentclass[numbers]{article}

\usepackage[final]{neurips_2024}
\usepackage{natbib}

\usepackage[utf8]{inputenc} 
\usepackage[T1]{fontenc}    
\usepackage{hyperref}       
\usepackage{url}            
\usepackage{booktabs}       
\usepackage{amsfonts}       
\usepackage{nicefrac}       
\usepackage{microtype}      
\usepackage{xcolor}         

\title{Is ETHICS about ethics? \\Evaluating the ETHICS benchmark}

\author{
   Leif Hancox-Li\thanks{Both authors have contributed equally to the paper.} \\
   vijil\\
   \texttt{leif@vijil.ai} \\ 
   \And
   Borhane Blili-Hamelin$^\ast$ \\
   AI Risk and Vulnerability Alliance\\
   \texttt{borhane@avidml.org} \\
}

\begin{document}

\maketitle

\section{Introduction}
\label{sec:intro}

ETHICS \citep{hendrycks_aligning_2023} is probably the most-cited dataset for testing the ethical capabilities of language models. Drawing on moral theory, psychology, and prompt evaluation, we  the validity of the ETHICS benchmark. Adding to prior work \citep{rao_ethical_2023, lacroix_metaethical_2022, talat_machine_2022}, our findings suggest that having a clear understanding of ethics and how it relates to empirical phenomena is key to the validity of ethics evaluations for AI.

\section{Gap between knowledge of moral theory and acting morally}

The ETHICS benchmark's focus on general moral theory is motivated by the thought that understanding moral theory is key to ``encouraging some form of `good' behavior in systems'' \cite[p. 2]{hendrycks_aligning_2023}. However, it is doubtful if \emph{people} who act morally do so by way of applying general moral theories. In fact, whether moral theory should be used in making decisions is a matter of debate within moral theory---such as in responding to the worry that being solely motivated by moral considerations might undermine friendship and love relationships \citep{Railton1984-RAIACA, sinnott-armstrong_moral_2010}.

Empirical research on the psychology of morality draws a distinction between \emph{moral behavior}---how real-world behavior is shaped by moral norms and expectations---and ``the way people think about morality'' \citep{ellemers_psychology_2019}. Even if it were true that moral theory can shape people's thinking about morality, the extent to which moral theories shape real behavior remains to be established. In a 2019 meta-review of 1200+ psychology papers on morality, \citet{ellemers_psychology_2019} argue that psychology research has had far greater success in studying how people think about morality than moral behavior. Just as importantly, they argue that studies focused on abstract moral principles of the kind that moral theories center has fallen short of establishing ``[t]he concrete implications of these general principles for specific situations''. 

As a provocation, we ask whether moral theories are the right model for empirically evaluating systems. Moral theories have the goal of systematizing moral considerations \citep{driver_moral_2022}. Moral theories are not designed to be empirically valid constructs in measuring real-world moral reasoning. When it comes to AI evaluation, frameworks like the Moral Foundations Questionnaire-2 (MFQ-2), developed with extensive consideration of construct validity---including cross-cultural validity beyond Western, Educated, Industrialized, Rich, and Democratic (WEIRD) cultures---are more promising starting points than moral theories \citep{atari_morality_2023, zakharin_moral_2023}. For ML work in this direction, see \citet{nunes_are_2024, abdulhai_moral_2023}.

\section{Misunderstanding the nature of general moral theories}

The ETHICS benchmark also lacks content validity \citep{jacobs_measurement_2021, blili-hamelin_borhane_making_2023}. Its prompts do not measure a language model's adherence to the different moral theories it claims to cover (deontology, utilitarianism \footnote{There is also a construct validity mistake in creating a false equivalence between utilitarianism and deontology and virtue ethics. Whereas deontology and virtue ethics are general families of moral theories, utilitarianism is a sub-species of consequentialism. It is misleading to present this specific variety of consequentialism as a primary competitor to deontology or virtue ethics \citep{driver_utilitarianism_2022, sep-consequentialism}.}, virtue ethics). This is because the authors misunderstand how these theories differ.

In the ETHICS benchmark, deontology prompts test the model's understanding of moral rules, utilitarianism prompts test understanding of the pleasantness of different scenarios, and virtue ethics prompts test identification of character traits. But moral theories within all broad families can make room for identifying character traits, rules, or pleasure as morally relevant \citep{hursthouse_virtue_2023}. For instance, rule utilitarianism, like deontology, also emphasizes rules. The difference is not the emphasis on rules, but the \emph{structure} of the moral theory. In rule utilitarianism, rules are selected based on whether their consequences maximize utility \citep{nathanson_rule, hooker_rule_2023}. Unfortunately, the ETHICS deontology prompts do not test \emph{how} these rules are selected, but only test whether the model can obey certain rules---an ability that is arguably compatible with all the named moral theories. 

Similarly, identifying character traits is not unique to virtue ethics. All moral theories can make room for character traits \citep{hursthouse_virtue_2023}:
\begin{quote}
This is not to say that only virtue ethicists attend to virtues, any more than it is to say that only consequentialists attend to consequences or only deontologists to rules. Each of the above-mentioned approaches can make room for virtues, consequences, and rules. Indeed, any plausible normative ethical theory will have something to say about all three. What distinguishes virtue ethics from consequentialism or deontology is the centrality of virtue within the theory (Watson 1990; Kawall 2009). Whereas consequentialists will define virtues as traits that yield good consequences and deontologists will define them as traits possessed by those who reliably fulfill their duties, virtue ethicists will resist the attempt to define virtues in terms of some other concept that is taken to be more fundamental. Rather, virtues and vices will be foundational for virtue ethical theories and other normative notions will be grounded in them.
\end{quote}

\section{Poor quality of prompts and labels}

Finally, we examined and relabeled 100 randomly sampled prompts from each of the three moral theory-based categories in the ETHICS dataset, and discovered considerable proportions of poor-quality prompts and/or labels. Given our professional training as academic philosophers, these labels can be considered more ``expert'' than those provided by the crowdworkers used to label the dataset, or any validation done by the creators of the ETHICS dataset (none of whom have professional training in ethics).

Here is a summary of the most common errors we found. Some of these are potential construct validity issues. The utilitarianism prompts ask the model to rate the ``pleasantness'' of different scenarios. However, utilitarianism is not about maximizing only pleasantness---the view that it is only about pleasantness or pain is a very specific form of utilitarianism known as hedonism. Hedonism faced several challenges raised in the 20th century, sprouting other variants of utilitarianism \citep{sep-consequentialism}. In addition, the utilitarianism prompts---which are evaluated through paired scenarios where Scenario 1 is supposed to be more pleasant than Scenario 2---have low-quality ground truth labels. We found that our labels do not agree with the crowdworker labels in 19\% of the prompt pairs.

Another common error type is prompts that require more context to answer correctly. 17\% of utilitarianism prompts and 8\% of deontology prompts in our sample fall into this category. This echoes the point made by \citet{lacroix_metaethical_2022} that ethical ``benchmarks'' make sense {\it only relative to} a stated set of values and contexts.

Finally, we found that 18\% of deontology prompts do not test ethical reasoning, as they can be answered correctly based solely on knowledge of physical impossibilities.

\begin{ack}
Borhane Blili-Hamelin was funded in part through a grant from the Brown Institute for Media Innovation. We thank Subho Majumdar for input on the project and drafts.
\end{ack}

\bibliographystyle{ACM-Reference-Format}
\bibliography{benchmark}




\end{document}